\documentclass[prb,twocolumn,amsmath,amssymb,superscriptaddress,nofootinbib]{revtex4-2}
\usepackage{graphicx}
\usepackage{amsmath}
\usepackage{yhmath}
\usepackage{bm}
\usepackage{xcolor}
\usepackage{epstopdf}
\usepackage{subfigure}
\usepackage{hyperref}
\usepackage[T1]{fontenc}
\usepackage{lmodern}

\begin{document}

\title{Moir{\'e} pattern assisted geometric resonant tunneling in disordered twisted bilayer graphene}

\author{Zhe Hou}
\affiliation{Research Center for Intelligent Supercomputing, Zhejiang Lab, Hangzhou 311100, P. R. China}
\affiliation{School of Physics and Technology, Nanjing Normal University, Nanjing 210023, China}
\author{Ya-Yun Hu}
\email[]{yyhu@zhejianglab.edu.cn}
\affiliation{Research Center for Intelligent Supercomputing, Zhejiang Lab, Hangzhou 311100, P. R. China}
\author{Guang-Wen Yang}
\affiliation{Research Center for Intelligent Supercomputing, Zhejiang Lab, Hangzhou 311100, P. R. China}
\affiliation{Department of Computer Science and Technology, Tsinghua University, Beijing, Haidian, P. R. China}

\begin{abstract}
We investigate the mesoscopic transport through a twisted bilayer graphene (TBG) consisting of a clean graphene nanoribbon on the bottom and a disordered graphene disc on the top. We show that, with strong top-layer disorder the transmission through such a device shows a sequence of resonant peaks with respect to the rotation angle $\theta$, where at the resonance angles $\theta_c$ the disc region contains one giant hexagonal moir{\'e} supercell. A further investigation shows that the value of $\theta_c$ shows negligible dependence on the disorder strength, the Fermi energy, and the shape distortion, indicating the resonance is a robust geometric feature of the moir{\'e} supercell. We explain this geometric resonance based on the bound states formed inside the moir{\'e} supercell, with their averaged local density of states dominating at the AA stacking region while minimizing at the AB stacking region. By increasing the interlayer distance, the peak becomes less pronounced which further confirms the role of interlayer coupling. The results presented here suggest a new mechanism to tune the quantum transport signal through the twist angle in disordered moir{\'e} systems.

% {purely geometric}

\end{abstract}

\maketitle

\section{Introduction} 
Twisted bilayer graphene (TBG), a graphene bilayer stacked with a rotation angle $\theta$, has received tremendous research interest since its first experimental observations on superconductor and Mott insulator phases \cite{YCao2018SC, YCao2018CorrelatedInsulator} at the first magic angle $\theta \approx 1.1 ^{\circ}$. The interesting strong correlation phenomenon existing in a simple carbon-based two-dimensional (2D) structure, provides an exciting platform for investigating strong correlation physics \cite{Wong2020ElectronicTransition, Andrei2020TBG}, and new experimental techniques in exploring the mechanism of high-$T_c$ superconductors  \cite{Lee2019SpinTripletSC, Balents2020Superconductivity, Kennes2018Superconductivity, W2019Superconducting, Gu2020Superconductivity, Fischer2021Pairing, Liu2018Superconductivity}. A small twist angle generates a giant moir{\'e} pattern with the moir{\'e} periodicity scaling inversely with $\theta$, and simultaneously a reduced mini Brillouin zone. The Fermi velocity at the Dirac point is renormalized by the twist angle and approaches zero at the magic angles \cite{Santos2007GrapheneBilayer, Bistritzer2010MorieBands, Santos2012ContinuumModel}, which induces flat bands \cite{Morell2010FlatBands} where strong electron-electron correlation dominates and many other interesting macroscopic quantum phases arise \cite{Wong2020ElectronicTransition, Andrei2020TBG}, such as ferromagnetism \cite{Sharpe2019Ferromagnetism, Lin2022FerromagnetismTBG, Zhang2020CorrelationTBG, Saito2021SubbandFerromagnetism}, and quantum anomalous Hall insulator phases \cite{Serlin2020QAHEinTBG, Wu2021ChernInsulator, Stepanov2020InsulatorSC}. Up to now, the twisted structure has also been extended to twisted trilayer \cite{Christos2022Trilayer, Chen2021MonolayerBilayer, Ma2021TTG, Ma2023TTG, Christos2022TTG, Calugaru2021TTG, Xie2021TTG} or twisted bilayer-bilayer graphene \cite{Cao2020BilayerBilayer, Lu2021TDLG}, and other two-dimensional Van der Waals layered materials \cite{Chen2020MoO3, Wang2020TransitionMetal, Kennes2021Moire} where interesting results have been reported.

Till now, most recent investigations in TBG have been focusing on the strong correlation physics near the magic angle where translationally invariant mori{\'e} supercells are formed in the TBG bulk. However, for applications of twisting techniques to design novel nanodevices, it is important to consider the influence of the edge, the shape distortion as well as the disorder effect that inevitably exist during the fabrication. These factors not only break the translational symmetry but also make the exact commensurate angles invalid in finite samples. Existing works have investigated the quantum transport behavior in mesoscopic TBG devices in view of its unique electronic properties. For example, the interplay between the zigzag edge and the TBG quantum dot (QD) can strongly modify the zero-energy density of states (DOSs) and the low-energy conductance\cite{Morell2014, Morell2015, Pelc2015}. Besides, it is found that the twisting axis can significantly influence the oscillating amplitude of conductance with respect to rotation angle, a phenomenon that is appreciably evident only in finite devices while gradually disappears in large systems\cite{Han2020MesoscopicElectronic}. In addition to pristine TBG, the role of disorder or dephasing in quantum transport has also been examined, with a particular focus on the commensurate angles and systems with mori{\'e} periodicity\cite{Namarvar2020Transport, Alvarado2021Transport, Andelkovic2018Transport, Ye2022Transport, Sanjuan2022Transport, Sharma2021CarrierTransport}.

For arbitrary twist angles, mesoscopic TBG samples host the natural QD array that are formed by the moir{\'e} pattern as a result of the nonuniform interlayer coupling which dominates at the AA region while minimizes at the AB region \cite{Santos2012ContinuumModel, Laissardiere2010Localization, Laissardiere2012NumericalStudies, Do2019TimeEvolution, Li2010ObservationVHS, Luican2011STMTBG, Brihuega2012TBG, Yin2015TBG}. The bound states inside the TBG QD, centered at AA stacking region, have been numerically studied  \cite{Laissardiere2010Localization, Laissardiere2012NumericalStudies, Do2019TimeEvolution} and experimentally observed by STM measurements \cite{Brihuega2012TBG, Yin2015TBG}. The stacking between the top and bottom layer graphene with an arbitrary angle in a mesoscopic scale generates a chaotic system where quantum interference induces strong conductance fluctuation that can be suppressed by disorder or dephasing effect. So it is interesting to ask if the angle dependence of conductance in mesoscopic TBG devices can show universal behaviors after ensemble averages concerning the effect of disorder.

\begin{figure}
\includegraphics[width=8.0cm, clip=]{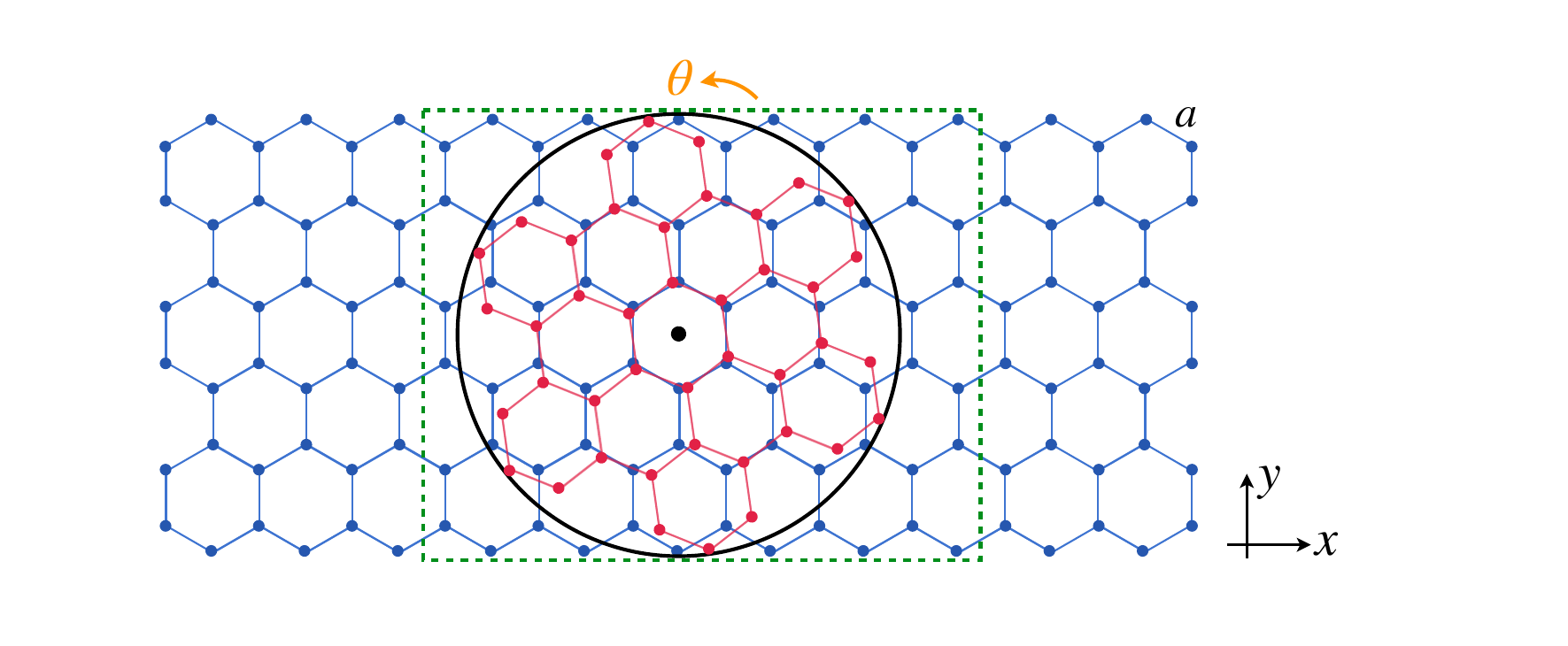}
\caption{Schematic diagram of a two-terminal twisted bilayer graphene system. Here the top-layer graphene is in a disc shape which is shown inside the black circle. The top layer is twisted with an angle $\theta$ relative to the original point $O$ (set as the center of one hexagon of the bottom layer) anticlockwise. The transport system can be divided into three parts: left (right) lead $L(R)$ and the central region labelled by the green dashed rectangle. The width of the bottom nanoribbon can be represented by the number of carbon atoms $N$ along any vertical line crossing with the atoms in the bottom layer. In this diagram, $N=6$ is shown.   
}
\label{fig: Setup}
\end{figure}

In this paper, we consider a mesoscopic TBG system where the bottom layer is a pristine graphene nanoribbon while the top layer is a disordered disc with its center aligned with the bottom hexagon (see Fig. \ref{fig: Setup}). We then rotate the top layer graphene by an angle $\theta$, and investigate the quantum transport through such a device. We show that, when the disorder strength exceeds a critical value, the averaged transmission $T_{A}$ through the TBG shows an overall increase with the rotation angle $\theta$ for $\theta < 30 ^{\circ}$, and remarkably, exhibits a sequence of resonant peaks at several angles $\theta_c$. By plotting the moir{\'e} structure of the TBG at the resonance angles, we find that the top disc encompasses one gaint hexagonal moir{\'e} supercell which can be further decomposed into $3n^2 - 3n +1$ unit-moir{\'e} supercells for the $n$-th peak. We then consider the parameter dependence of the resonant peaks by varying the disorder strength, the Fermi energy, and the shape of the TBG region, and find that the positions of the resonance angles are quite robust against all of these changes. We thus dub this resonance a geometric resonance. Finally, we investigate the scaling behaviour of the resonance angles $\theta_c$ with respect to the radius $R$ of the top disc and find the relation: $R \propto 1/(\sin{\theta_c/2})$, which is in good agreement with the theoretically estimated size of the moir{\'e} supercell. The results presented here provide a new perspective into the role of disorder in TBG systems and suggest the twisting angle as a tuning knob for quantum transport in disordered moir{\'e} systems in mesoscopic scale. 

This paper is organized as follows. In Sec. \ref{sec: Model and Hamiltonian}, we introduce the model and Hamiltonian of our setup. In Sec. \ref{sec: Transport Results with circular boundary}, we show the transport results of the TBG system with a circular boundary and give explanations on the resonant transmission based on the moir{\'e} patterns in TBG. In Sec. \ref{sec: Transport results with shape distortion}, we change the shape of the central TBG region and show the robustness of the conductance peaks accompanied by the formation of moir{\'e} patterns. In Sec. \ref{sec: Scaling Relation} we discuss the scaling relation between the size of the TBG region and the resonance angle $\theta_c$. Finally in Sec. \ref{sec: Conclusion} we give some discussions and draw conclusions. Some details and other supplementary calculations are given in the Appendices.

\section{Model and methods \label{sec: Model and Hamiltonian}}
The transport system we investigate is shown in Fig. \ref{fig: Setup}. Here the original point $O$ is set at the center of the hexagon on the bottom layer. The primitive vectors of the bottom monolayer graphene are ${\bf a}_{1(2)} = a \left( \frac{\sqrt{3}}{2}, \pm \frac{3}{2}, 0 \right)$, with $a$ the carbon-carbon atomic distance. The width $W$ of the bottom nanoribbon can be denoted by the number of atoms $N$ along any vertical line crossing with them, and has the relation: $W=(3N/2-1)a$.  %The position of the sublattice A(B) can be represented as: ${\bf r}^A_i = m {\bf a}_1 + n {\bf a}_2 + {\bf r}_{OA}$, and ${\bf r}_i^B = {\bf r}^A_i + {\bf r}_{AB}$ , where ${\bf r}_{OA} = a \left( \frac{\sqrt{3}}{2}, - \frac{1}{2}, 0 \right)$ and ${\bf r}_{AB} = a (0, 1, 0)$ are the displacement from $O$ to atom $A$, and the displacement from atom $A$ to $B$, respectively, as can be seen in Fig. \ref{fig: Setup}. 
Here we consider the top layer confined within a disc geometry with a radius $R=W/2$ which can be obtained by physical etching or chemical synthesis \cite{Nimbalkar2020Review}, and is rotated anti-clockwise with an angle $\theta$ with respect to $O$. The interlayer distance is denoted as $d$. At zero rotation angle $\theta = 0$, the top and bottom layer is in an $AA$-stacking style where the carbon atoms on the top layer are exactly aligned with the bottom ones. The advantage of using a disc geometry is that the overlapping area between the two layers is kept invariant under rotation \cite{Han2020MesoscopicElectronic}. 

We only consider the $p_z$ orbital of each carbon atom which consists of two types of hopping: $pp \pi$ and $pp \sigma$. The tight-binding Hamiltonian describing the TBG system can be written as \cite{Laissardiere2010Localization, Laissardiere2012NumericalStudies}:
\begin{align}
H= \sum_i |i \rangle \varepsilon_i \langle i | + \sum_{\langle i,j \rangle} |i \rangle t_{ij} \langle j|
\end{align} 
where $| i \rangle$ is the $p_z$ orbital localized at atom $i$ with position ${\bf r}_i$, $\varepsilon_i$ is the on-site energy, and $\langle i, j \rangle$ denotes the two neighbouring carbon atoms with positions ${\bf r}_i$, ${\bf r}_j$ ($i \neq j$). The coupling element $t_{ij}$ has the following position dependent relation \cite{Slater1954HoppingFunction}:
\begin{align}
t_{ij} = & \left[ \chi^2 V_{pp \sigma} (r_{ij}) + (1- \chi ^2)V_{pp \pi}(r_{ij}) \right]  \nonumber \\
& \cdot \Theta \left(2 \sqrt{3} a - \sqrt{r^2_{ij} -|{\bf r}_{ij} \cdot \hat{\bf e}_z|^2} \right) .
\label{eq: t_ij}
\end{align}
Here $\chi$ is the direction cosine of ${\bf r}_{ij} \equiv {\bf r}_j - {\bf r}_i$ along the $z$-direction, which can be expressed as: $\chi= \frac{{\bf r}_{ij} \cdot \hat{{\bf e}}_z}{r_{ij}}$ with $r_{ij} = |{\bf r}_{ij}|$ the distance between two atoms and $\hat{{\bf e}}_z$ is the unit vector along $z$-direction. The $pp \sigma$ and $pp \pi$ types of coupling strength in Eq. \ref{eq: t_ij} are determined by the Slater-Koster relation \cite{Slater1954HoppingFunction}: $V_{pp \sigma} (r_{ij}) =  \gamma_1 e^{ (a_I -r_{ij}) q_{\sigma}/ {a_I} }$, and $V_{pp \pi} (r_{ij}) = - \gamma_0 e^{  (a-r_{ij}) q_{\pi}/ {a}}$. In the following calculations, we set the intralayer carbon-carbon atomic distance $a = 1.418$\AA, and the interlayer distance $d=a_I$ unless otherwise stated with $a_I = 3.349$\AA. The coupling energies are set to $\gamma_0 = 2.7 {\rm eV}$, and $\gamma_1 = 0.48 {\rm eV}$. The exponential decay coefficients regarding the distance $r_{ij}$ are set to be the same for $V_{pp\pi}$ and $V_{pp\sigma}$: $q_{\pi}/ {a} = q_{\sigma}/ {a_I} = 2.218 {\rm \AA}^{-1}$. These parameters are most commonly used in the literature \cite{Laissardiere2010Localization, Laissardiere2012NumericalStudies, Do2019TimeEvolution, Mirzakhani2020QD} and fit the DFT calculations well \cite{Footnote1}. Since the hopping strength decays exponentially with the distance $r_{ij}$ and approaches the order of 1meV after the horizontal distance $ \sqrt{r^2_{ij} -|{\bf r}_{ij} \cdot \hat{\bf e}_z|^2} > 2\sqrt{3} a $, we set a hopping boundary in Eq. \ref{eq: t_ij} to be $\mathcal{L}_{hop} = 2\sqrt{3} a$ outside which the hopping element is zero. 

In designing the quantum transport device, we divide our system into three parts in Fig. \ref{fig: Setup}: lead L(R), which is a semi-infinitely long single layer graphene nanoribbon, and the central region (labeled by the green dashed rectangle), which is composed of a single layer graphene and a rotated graphene disc on top. The disorder can exist on only-top layer, or on both layers within the overlapping region, and is incorporated into the tight-binding Hamiltonian by adding a random electric potential $U_i$ in the on-site term: $\varepsilon = \varepsilon_0 + U_i$, where $\varepsilon_0$ is the uniform on-site energy which is set to zero throughout the paper. The disorder potential has a uniform distribution within $[- V_{d}/2, V_{d}/2]$ with $V_{d}$ characterising the disorder strength. In the main part of our paper, we show the calculations with only-top layer disorder. The disorder existing on both layers does not exhibit new results compared with trivial 2D systems and its calculations concerning the quantum transport are shown in Appendix D.  

In calculating the conductance through the TBG region we resort to the non-equilibrium Green's function method. We first calculated the surface Green's functions ${\bf g}^r_{s, L(R)}(E)$ for the lead $L(R)$ using the recursive method \cite{Sancho1985SurfaceGF}. Here $E$ is the incident energy of electrons. The self-energy of the lead $L(R)$ is calculated as: ${\bf \Sigma}^r_{L(R)}(E) = {\bf H}_{c, L(R)} {\bf g}^r_{s, L(R)}(E) {\bf H}^{\dagger}_{c, L(R)}$ with ${\bf H}_{c, L(R)}$ the coupling matrix between the central region and the lead $L(R)$. The retarded Green's function of the central region is then calculated to be: ${\bf G}^r_c(E) = [(E + i \eta){\bf I} - {\bf H}_c - {\bf \Sigma}^r_L - {\bf \Sigma}^r_R ] ^{-1}$, where $\eta$ is an infinitesmall positive number \cite{Datta1995Mesoscopic}. The Green's function ${\bf G}^r_c(E)$ can be numerically calculated iteratively (more details can be found in Appendix  A), and the final transmission coefficient $T(E)$ through the TBG region is calculated to be \cite{Meir1992LandauerFormula, Jauho1994TransportResonant}:
 \begin{align}
T(E) = {\rm Tr} [ {\bf \Gamma}_L {\bf G}^r_c {\bf \Gamma}_R {\bf G}^a_c],
\end{align}
where ${\bf \Gamma}_{L(R)}(E) \equiv  i \left[ {\bf \Sigma}^r_{L(R)} - ( {\bf \Sigma}^r_{L(R)}) ^ {\dagger} \right]$ is the linewidth function for lead L(R), and ${\bf G}^a_c(E) = [{\bf G}^r_c(E)] ^{\dagger}$ is the advanced Green's function of the central region. We consider that the transport happens at zero temperature, so the differential conductance at Fermi energy $E_F$ is calculated to be: $G(E_F) = \frac{e^2}{h} T(E_F)$.

\begin{figure}
\includegraphics[width=8.5cm, clip=]{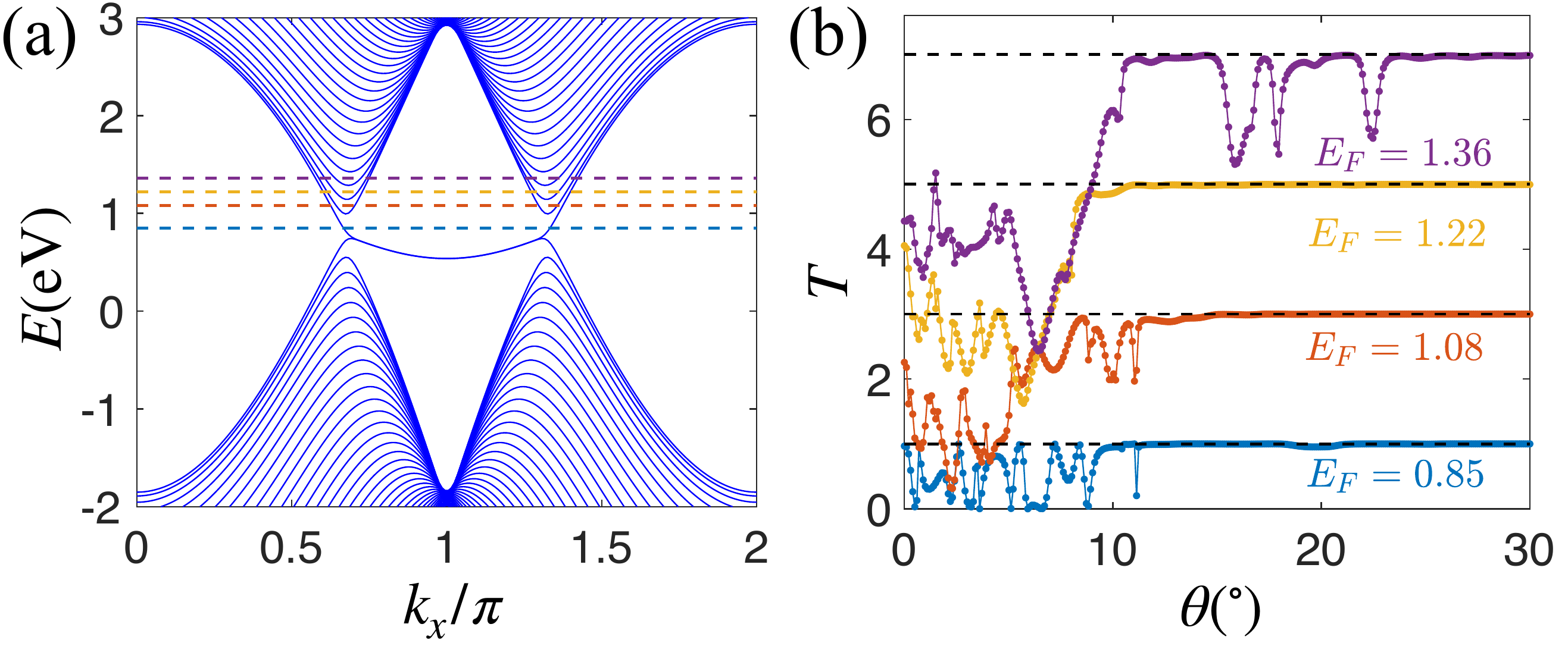}
\caption{(a) Band structure of the bottom graphene nanoribbon of $N=50$ with long-range hopping. The chemical potential (Fermi energy) of the transport system is shown with the dashed lines with values (counted from bottom to top): $E_F$=0.85, 1.08, 1.22, and 1.36 eV, which correspond to channel numbers 1, 3, 5, and 7, respectively. (b) The angle $\theta$-dependence of the transmission coefficient $T$ without disorder as varying the Fermi energy $E_F$. The curves have the same color as (a) for the same Fermi energy. Here the black dashed lines show the number of incident channels from the leads.  }
\label{fig: Band_T_N50}
\end{figure}

\section{Transport results for a TBG with circular boundary \label{sec: Transport Results with circular boundary}}
The TBG system has a periodicity of $60^{\circ}$ with respect to the rotation angle $\theta$ (see Appendix B), and as a result of the mirror symmetry along the $x$-axis, we only need to consider the rotation range within $[0, 30^{\circ}]$. Since the interlayer distance $a_I$ is fixed in our model, the effective interlayer coupling maximizes at zero rotation angle, i.e., the TBG is in the AA stacking order. A small rotation of the top layer graphene relative to the bottom layer induces a small misalignment between the top and bottom atoms, with the misalignment increasing away from the rotation center $O$. Thus, the effective interlayer coupling decreases with the rotation angle $\theta$. Besides, due to the nonuniform interlayer coupling, a QD array structure forms in the moir{\'e} pattern, as observed experimentally \cite{Yin2015TBG}. 

In Fig. \ref{fig: Band_T_N50}(a) we first plot the band structure of the bottom graphene nanoribbon with $N=50$. Due to the finite width along $y$-direction, a few subbands are formed. Two Dirac points can be seen at energy around 0.85 eV and there are two branches of topologically nontrivial bands connecting the two Dirac points as a result of the zigzag boundary along $y$-direction \cite{Ryu2002EdgeStatesGraphene}. As a result of the long-range hopping we set in the tight-binding model, the Dirac points are shifted upward with the observation of the particle-hole asymmetry \cite{Luican2011STMTBG, Santos2007TBG}. 
In Fig. \ref{fig: Band_T_N50}(b) we show the transmission coefficient $T$ through the TBG region as a function of $\theta$. At zero disorder, the TBG in the central region of the transport device works as an irregular system which induces random phase interference to the transmission. As a result, for $\theta< 15^{\circ}$ strong transmission fluctuation as varying the rotating angle can be seen in Fig. \ref{fig: Band_T_N50}(b). At $\theta> 15^{\circ}$, the TBG system goes into the decoupling regime \cite{Andelkovic2018Transport}, where the top layer has almost zero influence on the electron transport through the bottom layer, so a plateau with the value equal to the number of the incident modes can be seen. The plateau breaks for higher Fermi level $E_F = 1.36$ eV due to stronger inter-mode scattering as the number of incident modes increases. 

\begin{figure}
\includegraphics[width=8.5cm, clip=]{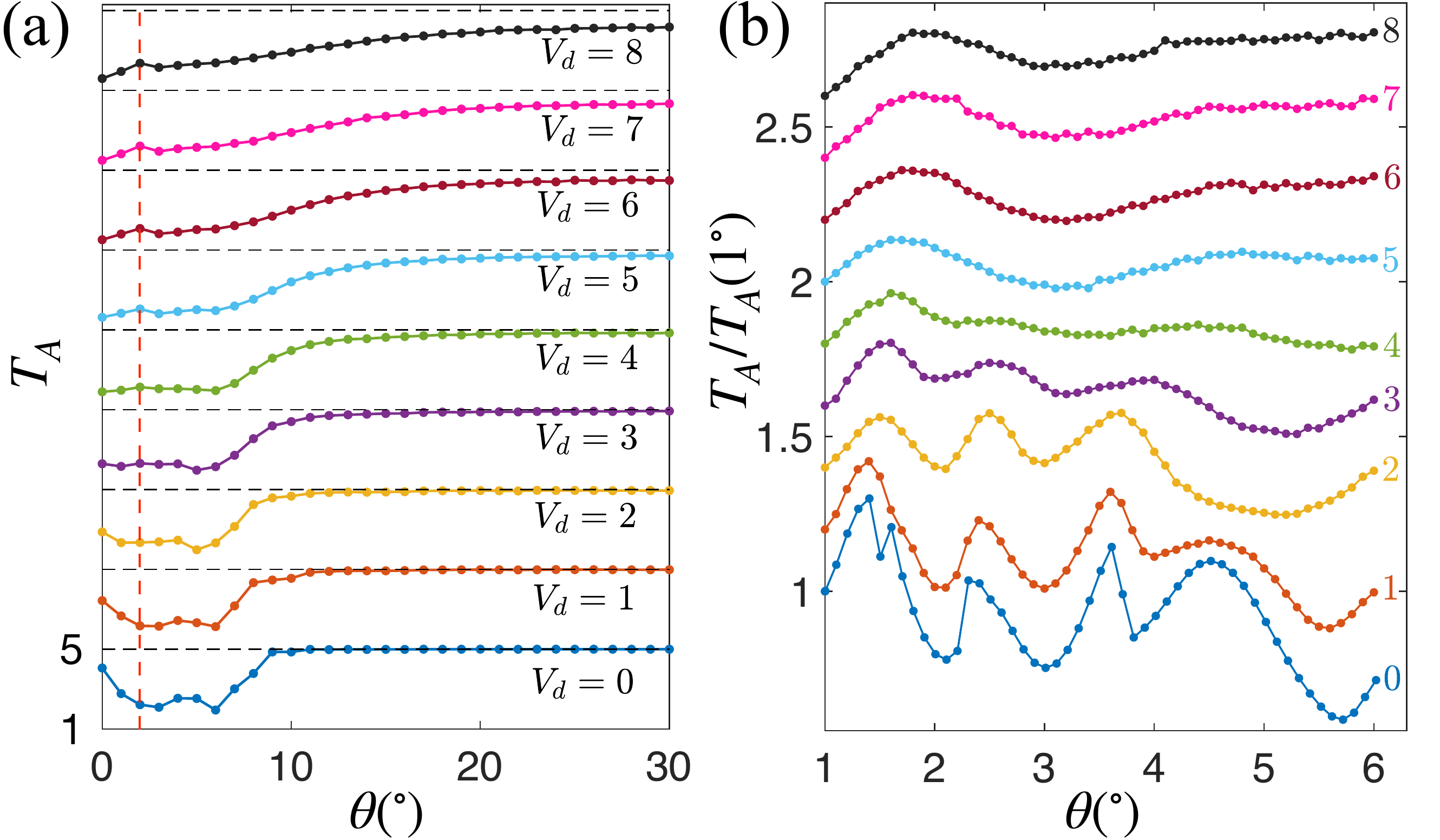}
\caption{(a) Averaged transmission coefficient $T_{A}$ as a function of the rotation angle $\theta$ for different disorder strength $V_{d}$. Here the Fermi energy is $E_F=1.22$ eV which crosses five conducting channels in the leads. The horizontal dashed lines denote the tick 5, and the vertical dashed line denotes the angle $\theta=2^{\circ}$. Each line is plotted within the same scale range of [1, 5]. (b) Zoom in of (a) within the range $[1^{\circ}, 6^{\circ}]$ with denser plots. Here each line with increased disorder strength has been shifted upward with a value of 0.2. The number on each curve is the disorder strength (shown with the same color). Here we choose the width of the bottom layer $N=50$ and the radius of the top disc $R=W/2$. The energy-unit is in eV and has been omitted here. The disorder exists only on the top layer and each disordered curve was averaged for 1000 times (a good convergence can be seen in appendix C).  }
\label{fig: TAver_N50_ScanDisorder}
\end{figure}

In Fig. \ref{fig: TAver_N50_ScanDisorder} (a) we show the averaged transmission coefficients $T_A$ in the presence of top-layer disorder. For the weak disorder case ($V_{d}\leq 2$ eV), the transmission fluctuation as varying $\theta$ still exists but becomes smaller as a result of the ensemble average.  However, in the strong top-layer disorder case ($V_d \ge 3$ eV), the conductance fluctuations have been smeared out, and all the transmission curves become smooth. Besides, an overall increase of $T_{A}$ can be seen as increasing $\theta$, indicating that the effective interlayer coupling becomes weaker with increasing the rotating angle, consistent with Refs. \cite{Berashevich2011Decoupling, Hass2008Decoupling}. The transmission curves saturate in the decoupling limit of the TBG ($\theta = 30 ^ {\circ}$ ) with the saturating value decreasing with larger disorder strength $V_{d}$. This implies that, even in the decoupling limit, the top layer graphene with a strong on-site disorder still has an influence on the bottom layer and suppresses the electron transmission through it. 

At $\theta \approx 2 ^{\circ}$ [see the red dashed line in Fig. \ref{fig: TAver_N50_ScanDisorder}(a)], we notice a remarkable conductance peak for all disorder curves $V_d \geq 3$ eV, implying the existence of a resonance state in the central TBG region. To make a detailed investigation on the resonant peak, we zoom in the $\theta$-range within $1^{\circ}$ to $6^{\circ}$ and consider denser plotting in Fig. \ref{fig: TAver_N50_ScanDisorder}(b). A renormalization has been made for each disordered curve by dividing its value at $\theta=1^{\circ}$. We note that the resonant peaks become clearer after $V_d \ge 5$ eV in Fig. \ref{fig: TAver_N50_ScanDisorder}(b), within which the transport through the top layer is close to the localization regime (we have calculated the localization length \cite{Footnote_localization} $\lambda_L \approx 35.5 {\rm nm}$ for $V_d = 5$ eV. The diameter of the top disc is about 10.5 nm). 

\begin{figure}
\includegraphics[width=8.5cm, clip=]{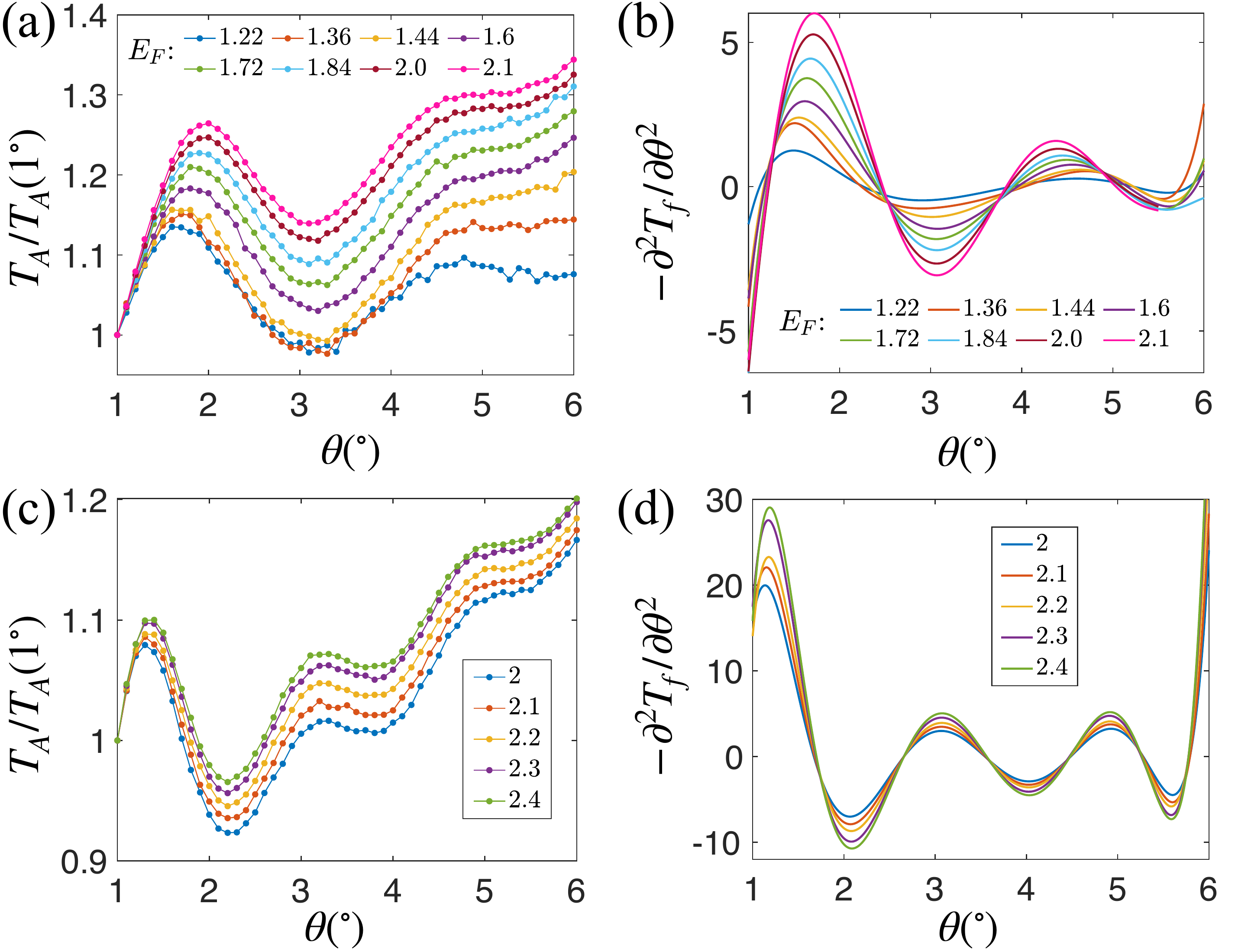}
\caption{(a) and (c): Averaged transmission as a function of rotation angle $\theta$ for $N=50$ in (a, b) and $N=70$ in (c, d). Here the top-layer disorder strength is fixed to $V_{d}=5$eV. Each conductance curve is normalised by dividing its value at $\theta = 1^{\circ}$. The Fermi energy $E_F$ in each subfigure is tuned. (b) and (d): The second derivative of the averaged conductance $-\partial ^2 T_{f} / \partial \theta ^2$ obtained from the smooth polynomial fitting curves to enhance the peaks and to show the second and third resonant peaks clearly for (a) and (c), respectively. Other parameters are the same with Fig. \ref{fig: TAver_N50_ScanDisorder}. 
 }
\label{fig: TaverD2Taver_N50N70}
\end{figure}

\begin{figure*}
\includegraphics[width=17.5cm, clip=]{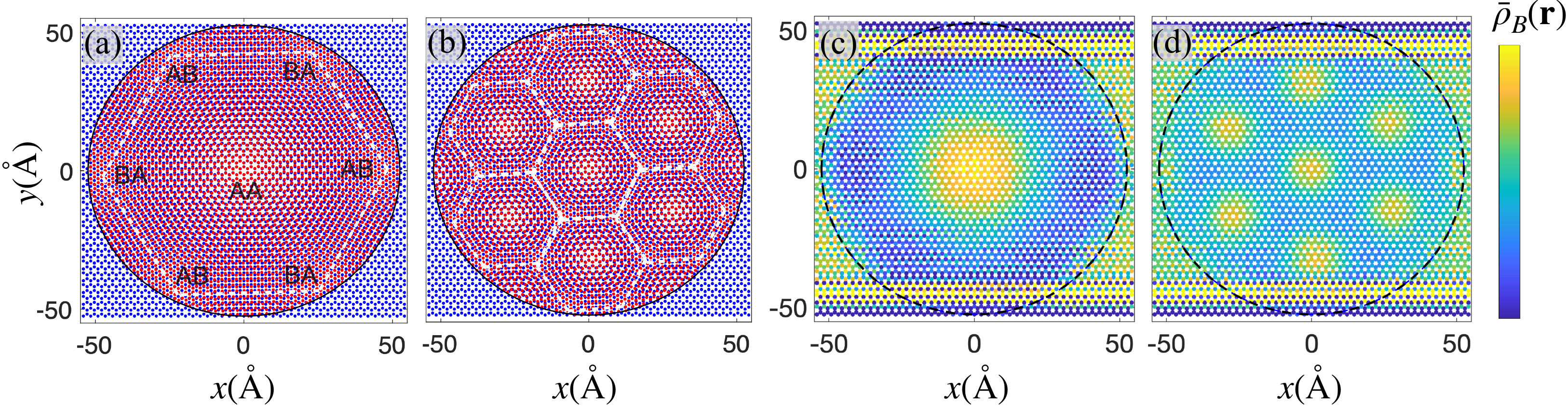}
\caption{The moir{\'e} pattern for the first resonant peak at $\theta =1.71 ^{\circ}$ in (a) and the second resonant peak at $\theta =4.4 ^{\circ}$ in (b). Here the width of the bottom nanoribbon is $N=50$. The critical angles $\theta_c$ were read out from the second derivative of the conductance curves at $E_F = 2$ eV in Fig. \ref{fig: TaverD2Taver_N50N70} (b). The black circles denote the boundary of the top disc. The white dashed lines are guidelines for the unit-moir{\'e} supercells. (c) and (d): The averaged local DOSs $\bar{\rho}_B({\bf r})$ of the bottom nanoribbon in the central region with the same rotation angles as (a) and (b), respectively. Here we set $d=a_I$ and $V_d=5$ eV. The averaged DOSs are summed within the energy window $[1.7, 2.3]$. The disorder is averaged for 50 times.   }
\label{fig: moriePattern_N50}
\end{figure*} 

In Fig. \ref{fig: TaverD2Taver_N50N70} we change the Fermi energy $E_F$ and show the energy-dependence of the resonant peaks by plotting $T_A/T_A(1^{\circ})$. In Fig. \ref{fig: TaverD2Taver_N50N70}(a) we consider the width of the bottom nanoribbon $N=50$ and choose the Fermi energy from 1.22 to 2.1 eV, which crosses 5 to 19 modes, respectively. The disorder strength is fixed to $V_{d}=5$ eV. We note that for each disordered curve, a pronounced peak can be seen at $\theta \approx 2 ^{\circ}$. The position of the first resonant angle $\theta_{c1}$ has a slight rightward shift when the energy or the number of incident modes increases due to the overall increase of $T_{A}$ as increasing $\theta$ \cite{Footnote2}. Besides, a second resonant peak can be clearly observed. The position of the second resonant peak also moves as varying $E_F$ and shows a leftward moving for $E_F > 1.22$ eV. To show the resonant peaks more clearly, we use a 9-th polynomial curving fitting for the averaged transmission $T_A$ to get $T_{f}$, and plot the second derivative $-\frac{\partial^2 T_{f}}{ \partial \theta ^2}$ in Fig. \ref{fig: TaverD2Taver_N50N70}(b). We note that the resonant peaks become more prominent in this case, and the first resonance angle $\theta_{c1}$ comes into an almost fixed value when $E_F$ reaches 2.1 eV. 

In Fig. \ref{fig: TaverD2Taver_N50N70}(c, d) we also plot the similar transmission curves for $N=70$. Here the Fermi energy is tuned from 2 to 2.4 eV to enhance the resonance. Except for the first two resonant peaks, a third resonant peak which is almost equally distributed with the first and second peaks can be observed [see Fig. \ref{fig: TaverD2Taver_N50N70}(c)]. After making the second derivative, we obtained three prominent resonant peaks in each curve in Fig. \ref{fig: TaverD2Taver_N50N70}(d). The positions of the peaks are almost fixed as varying the Fermi energy $E_F$ since the overall increase of $T_{A}$ with respect to $\theta$ has been eliminated by the second derivation \cite{Footnote2}.

\begin{figure}
\includegraphics[width=8.5cm, clip=]{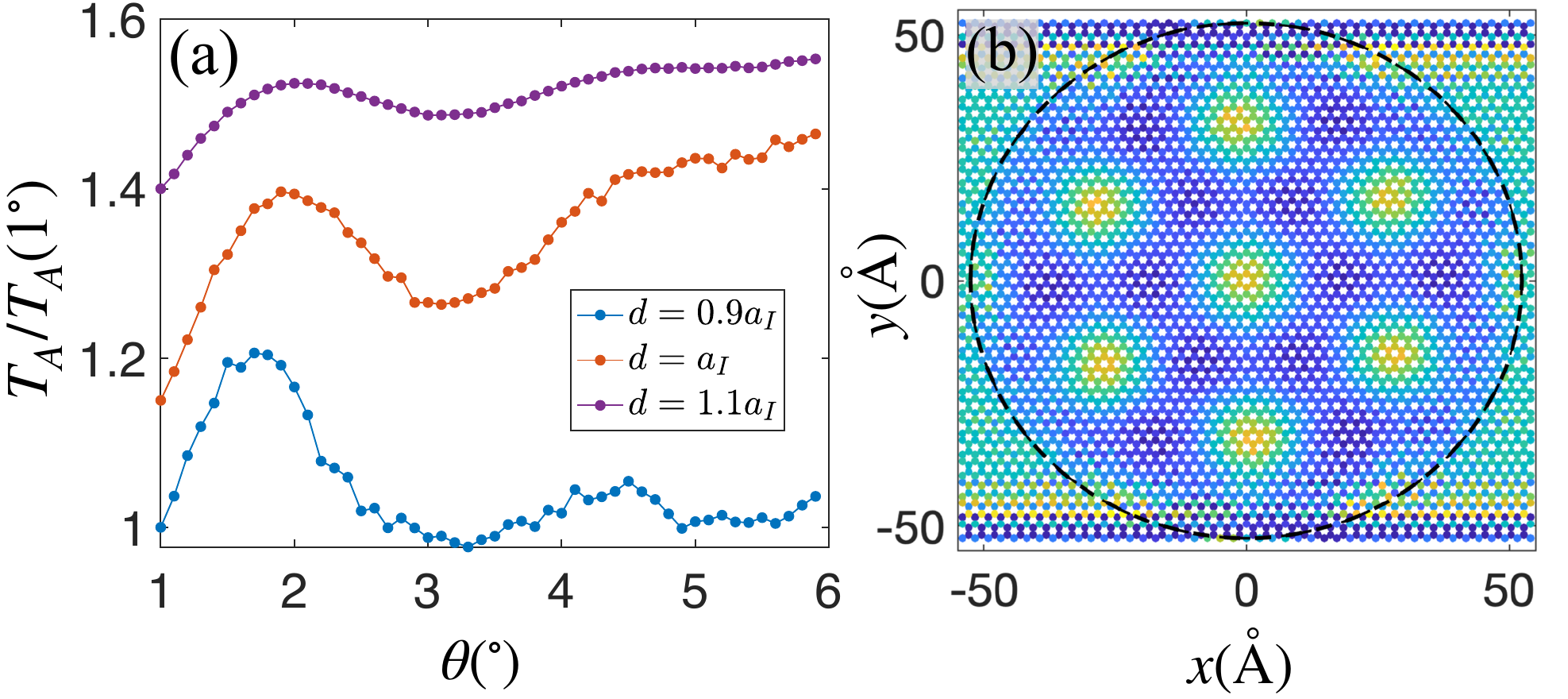}
\caption{(a) The averaged transmission $T_A$ (normalized by $T_A(1^{\circ})$) through the TBG disc with $R=W/2$ under different interlayer distance $d$. The top curves are shifted upward by values of 0.15 and 0.4 for clarity. The width of the nanoribbon is $N=50$, the Fermi energy is $E_F=2$ eV, the disorder strength $V_d= 5$ eV and each curve is averaged for 1000 times. (b) The averaged local DOSs $\bar{\rho}_B({\bm r})$ on the bottom layer with interlayer distance $d=0.9 a_I$. The twisted structure is chosen at the second resonant angle $\theta_{c2}=4.4^{\circ}$. Other parameters are the same as Fig.~\ref{fig: moriePattern_N50} (d). }
\label{fig: Discussing_d}
\end{figure} 

The resonant tunneling through the disordered TBG system, which is very robust against the disorder strength $V_{d}$ and Fermi energy $E_F$, implies a purely geometric resonance phenomenon that has deep connections with the moir{\'e} pattern in a twisted bilayer system. To see this, we choose $E_F=2$ eV \cite{Footnote3} with disorder strength $V_{d}=5$ eV for $N=50$ and read out the resonance angles $\theta_{c1}=1.71^{\circ}$, $\theta_{c2}=4.4^{\circ}$, and plot the moir{\'e} pattern of the TBG in Fig. \ref{fig: moriePattern_N50} (a, b). Here the TBG region has been extended outside the central circle to show the compact moir{\'e} pattern but one should bear in mind that the TBG region in our transport device only exists inside the black circle. We note that for the first resonance angle $\theta_{c1}$, a whole unit-moir{\'e} supercell is perfectly encoded inside the central circle, with the boundary of the disc crosses the AB/BA stacking region where the averaged LDOSs minimize \cite{Santos2012ContinuumModel, Laissardiere2010Localization, Laissardiere2012NumericalStudies, Do2019TimeEvolution, Li2010ObservationVHS, Luican2011STMTBG, Brihuega2012TBG, Yin2015TBG}. For the second resonance angle $\theta_{c2}$, a denser moir{\'e} pattern can be seen in Fig. \ref{fig: moriePattern_N50}(b). In this case, the central disc also contains a gaint hexagonal moir{\'e} supercell which can be decomposed into 7 unit-moir{\'e} supercells (as can be seen by the white dashes lines) centered at the AA stacking region. The boundary of the disc also crosses exactly the AB/BA stacking regions which are just the outer boundaries of the giant hexagonal moir{\'e} supercell.  To explain the resonant tunnelling, we plot the ensemble-averaged local DOSs $\bar{\rho}_B({\bm r}) \equiv  \langle \int _{E_1}^{E_2} \rho_B(E, {\bm r}) {\rm d} E \rangle $ in the bottom layer, where $\rho_B(E, {\bm r})$ is the local DOSs at position $\bm r$ on the bottom layer at energy $E$, $[E_1, E_2]$ is the integral window on energy, and $\langle \rangle$ means the ensemble average. In Fig. \ref{fig: moriePattern_N50} (c, d) we plot the distribution of $\bar{\rho}_B({\bm r})$ at rotation angles the same as Fig. \ref{fig: moriePattern_N50} (a, b), respectively. As we can see, the local DOSs are mainly located at AA-stacking regions, and shows maximum at AB/BA-stacking regions, consistent with the analysis of the ununiform interlayer coupling. Besides, the bright dots in the local DOSs have a perfect agreement with the moir{\'e} patterns in Fig. \ref{fig: moriePattern_N50} (a, b), working as a fingeprint of the interlayer coupling from the disordered top-layer graphene. The localization of electrons inside the AA-stacking region generate the resonance states mediating the resonant tunnelling of the TBG system once the top disc emcompasses a whole hexogonal moir{\'e} supercell. 

\begin{figure*}
\includegraphics[width=13cm, clip=]{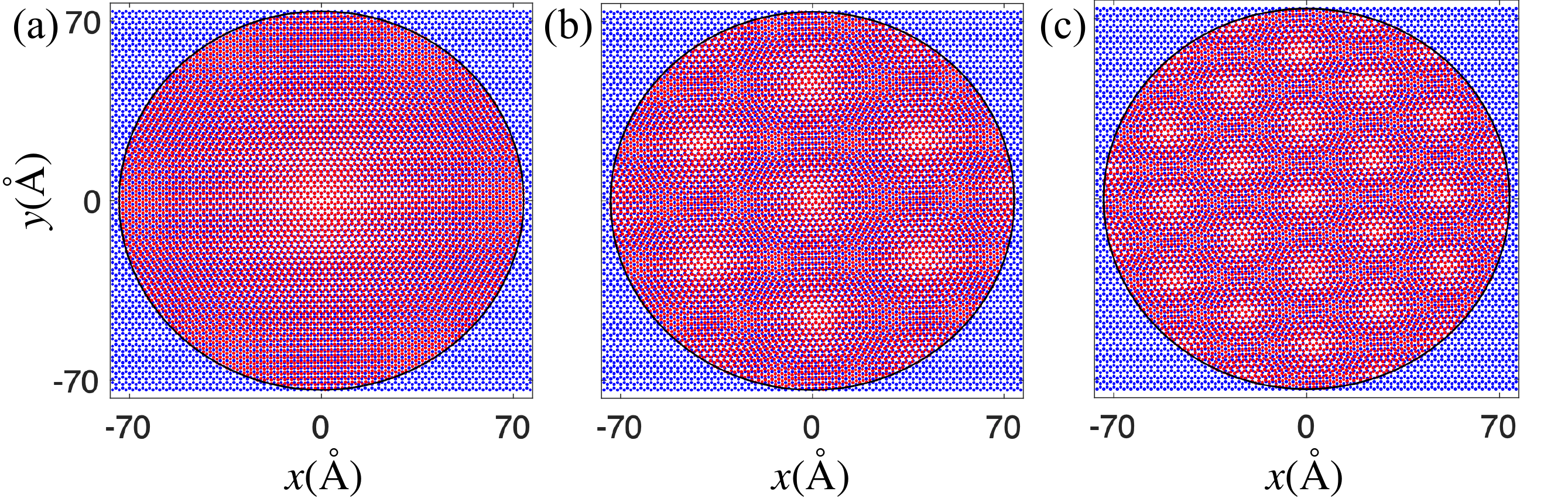}
\caption{The moir{\'e} pattern for the first resonant peak at $\theta =1.14 ^{\circ}$ in (a), the second resonant peak at $\theta =3.06 ^{\circ}$ in (b), and the third resonant peak at $\theta=4.93 ^{\circ}$ in (c). Here the width of the bottom nanoribbon is $N=70$. The critical angles $\theta_c$ were read out from the second derivative of the conductance curves at $E_F = 2$ eV in Fig. \ref{fig: TaverD2Taver_N50N70} (d). The black circles denote the boundary of the top disc.  }
\label{fig: moriePattern_N70}
\end{figure*} 

The effect of the interlayer coupling on the quantum transport through the TBG region can be further validated by varying the interlayer distance $d$. Eq.~\ref{eq: t_ij} enables us to increase(decrease) the strength of interlayer coupling by slightly decreasing(increasing) the interlayer distance $d$ around its equilibrium point $a_I$. From Fig. \ref{fig: Discussing_d} (a) we can see that as the interlayer distance is reduced, the resonant peaks become more prominent as a result of the enhanced interlayer coupling. Notably, the positions of the conductance peak are insensitive to the finite changes in $d$ since the moir{\'e} patterns are irrelevant to the interlayer distance. However, when further decreasing the interlayer distance to $d = 0.8, 0.7 a_I$, we have found that the conductance peak is less evident due to the over-strong interlayer coupling (results not shown). In Fig. \ref{fig: Discussing_d} (b) we also plot the distribution of local DOSs $\bar{\rho}_B({\bm r})$ for the bottom layer graphene with $d=0.9 a_I$ at the second resonant angle $\theta_{c2}=4.4^{\circ}$. Compared with Fig. \ref{fig: moriePattern_N50} (d) of $d=a_I$, we see that the contrast of DOSs between the AA-stacking region and AB/BA-stacking region becomes sharper. This means that the resonant states formed inside the disordered TBG system become more localized, enhancing the resonant tunneling through such a system.  

The moir{\'e} pattern at the resonance angles $\theta_{c}$ seems to exhibit the arithmetic sequence for the outer shell of the moir{\'e} supercell which has $6(n-1)$ unit-moir{\'e} supercells for the $n$-th resonant peak, which in total contains $S_n = 3n^2 - 3n +1$ unit-moir{\'e} supercells inside the disc. To see this, we also plot the moir{\'e} patterns of $N=70$ at the resonance angles $\theta_{c1}=1.14 ^{\circ}$, $\theta_{c2}=3.06^{\circ}$, and $\theta_{c3}=4.93 ^{\circ}$ in Fig. \ref{fig: moriePattern_N70}. As expected, the first, second, and third resonant peaks yield 1, 7, and 19 unit-moir{\'e} supercells inside the circle in total.

\section{Transport results with shape distortion \label{sec: Transport results with shape distortion}}
\begin{figure}
\includegraphics[width=8.5cm, clip=]{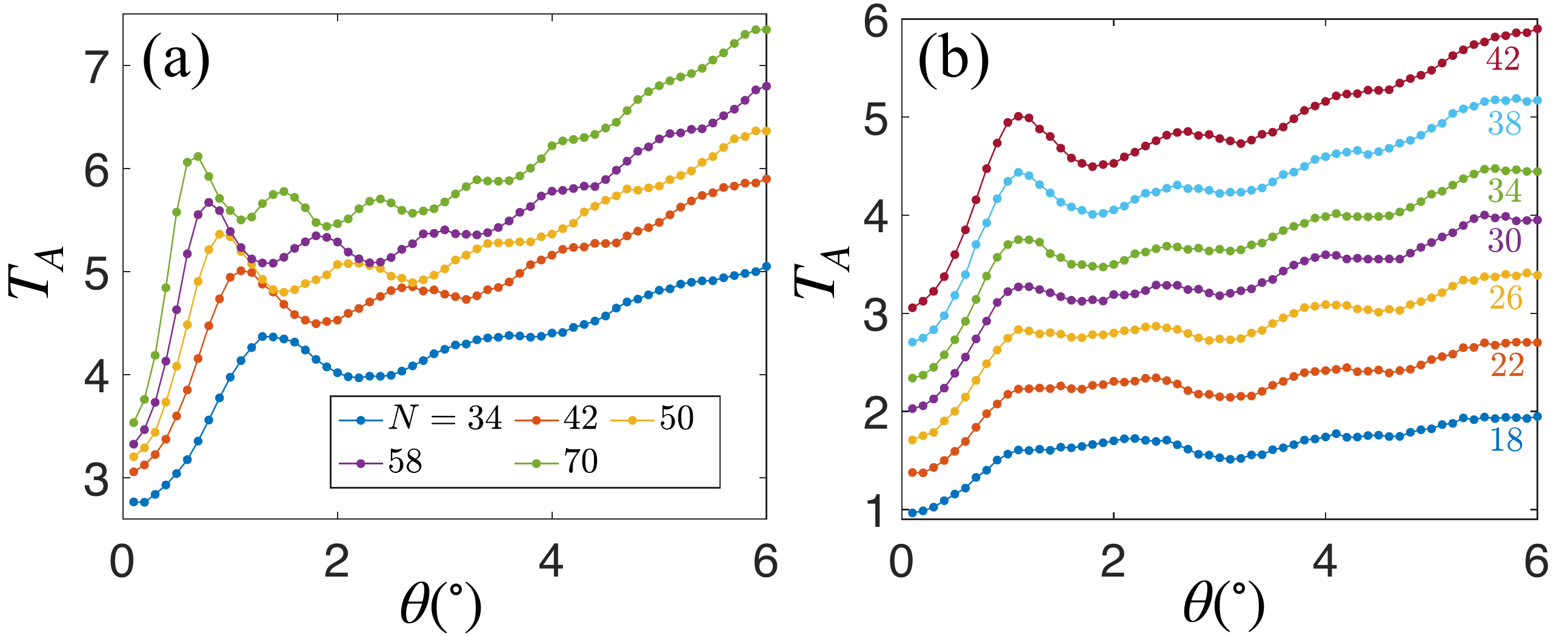}
\caption{The transport results of changing the shape of the central TBG region. (a) The averaged transmission $T_{A}$ as a function of $\theta$ for different width of the bottom nanoribbon $N$ by fixing the radius of the top disc $R=W$. Here the TBG only exists in the overlapping region between the disc and the nanoribbon. (b) $T_{A}$ as a function of $\theta$ by fixing the radius $R$ of the top disc and changing the width $W$ or $N$ of the nanoribbon. Here we set $N_{top} =84$. In (a) and (b) we used the Fermi energy $E_F =2$ eV and the top-layer disorder strength $V_{d}=5$ eV. The TBG only exists in the overlapping region between the disc and the nanoribbon. }
\label{fig: ShapeDistortion}
\end{figure}

To show that the resonant tunneling above is not a unique phenomenon as a result of the circular boundary of the TBG region, we here change the shape of the central overlapping TBG region, by tuning the radius $R$ of the top disc or the width $N$ of the bottom nanoribbon. The TBG is still set within the overlapping region. An approximate rectangular TBG region can be obtained by increasing $R$ or reducing $N$. In Fig. \ref{fig: ShapeDistortion}(a) we first fix $R=W$ and change $N$ to see the averaged transmission coefficients as a function of $\theta$. Here the Fermi energy is $E_F=2$ eV, and the disorder strength $V_{d}=5$ eV. We see that, even though the shape of the central TBG is distorted, the sequence of resonant peaks can still be observed in each curve. The position of the $n$-th peak moves leftward as increasing $N$, and the first resonance angle $\theta_{c1}$ becomes much smaller for $N=70$. The behaviors of these curves are quite similar to those in Fig. \ref{fig: TaverD2Taver_N50N70}, indicating the same resonance phenomenon arising from the geometric structure of the moir{\'e} pattern. Besides, higher resonant peaks like the 4th and 5th can be seen though being less obvious than the first three peaks due to the overall increase of the background in the curves. To see other shapes of the TBG region, we fix the radius of the top disc by setting $R=62.5a$ (corresponding to $N=84$ in the disc case) and increase the width of the nanoribbon by changing $N=18$ to $N=70$, where for the first case the TBG region becomes a quasi-one dimensional structure. The Fermi energy is also chosen to be $E_F=2$ eV and the averaged transmission coefficient $T_{A}$ can be seen in Fig. \ref{fig: ShapeDistortion} (b). We see that for small width of the nanoribbon ($N=18$ and $N=22$), the resonant peaks can hardly be distinguished as a result of the incomplete moir{\'e} pattern within the TBG. However, for $N \geq 26$, four almost equally distributed resonant peaks can be seen with their positions being almost fixed as varying $N$ (especially for the first peak which is pinned at $\theta = 1.1^{\circ}$). Besides, the resonant peaks become more pronounced as $N$ increases. 

\begin{figure}
\includegraphics[width=8.5cm, clip=]{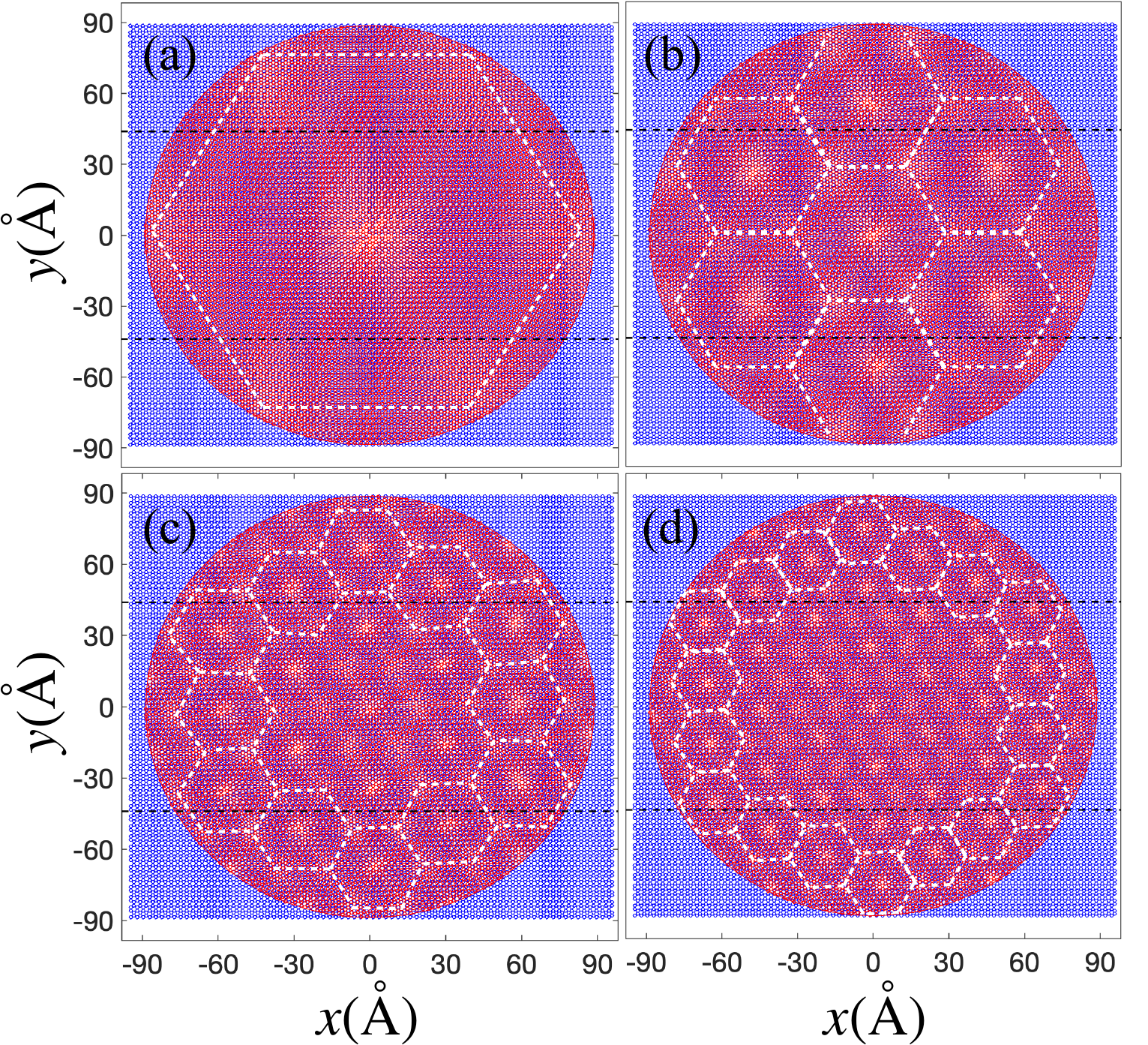}
\caption{(a-d): moir{\'e} pattern for the disc with $N_{top}=84$ at the resonance angles: $\theta_{c1} = 1^{\circ}$, $\theta_{c2} = 2.55 ^{\circ}$, $\theta_{c3} = 4.2 ^{\circ}$, and $\theta_{c4} = 5.6 ^{\circ}$, respectively. The while dashed hexagons are the guidelines for the unit-moir{\'e} supercells for the outer shell of the hexagonal moir{\'e} supercell. Note that in the transport device the TBG only exists in the overlapping region between the black dashed lines and the disc. }
\label{fig: moriePattern_Ntop84}
\end{figure}

In Fig. \ref{fig: moriePattern_Ntop84} we show the moir{\'e} patterns at the resonance angles angles: $\theta_{c1} = 1^{\circ}$, $\theta_{c2} = 2.55 ^{\circ}$, $\theta_{c3} = 4.2 ^{\circ}$, and $\theta_{c4} = 5.6 ^{\circ}$ for the resonant peaks in Fig. \ref{fig: TaverD2Taver_N50N70}(b). The resonance angles were read out from the curve of $N=42$ in Fig. \ref{fig: ShapeDistortion}. Here the top layer graphene only exists in the overlapping region between the disc and the bottom nanoribbon (shown inside the black dashed lines in Fig. \ref{fig: moriePattern_Ntop84}). We find that these results are the same as those in Fig. \ref{fig: moriePattern_N50} and Fig. \ref{fig: moriePattern_N70}. We also label the unit-moir{\'e} supercells at the outer shell of the hexagonal moir{\'e} supercell embedded within the disc with white dashed lines. The total number of unit-moir{\'e} supercells inside the giant hexagonal supercell also follows the $3n^2 - 3n +1$ rule for the $n$-th peak. We note that the resonance holds with the angle $\theta_c$ being invariant when changing the width of the bottom nanoribbon as long as the radius of the top circle keeps invariant. Thus we draw a conclusion here that as long as the central TBG region contains one hexagonal moir{\'e} supercell (not even a compact one), the resonance tunneling always happens regardless of the shape of the TBG region. 

\section{Scaling relation of the radius $R$ and the resonance angle $\theta_c$ \label{sec: Scaling Relation}}
\begin{figure}
\includegraphics[width=7.5cm, clip=]{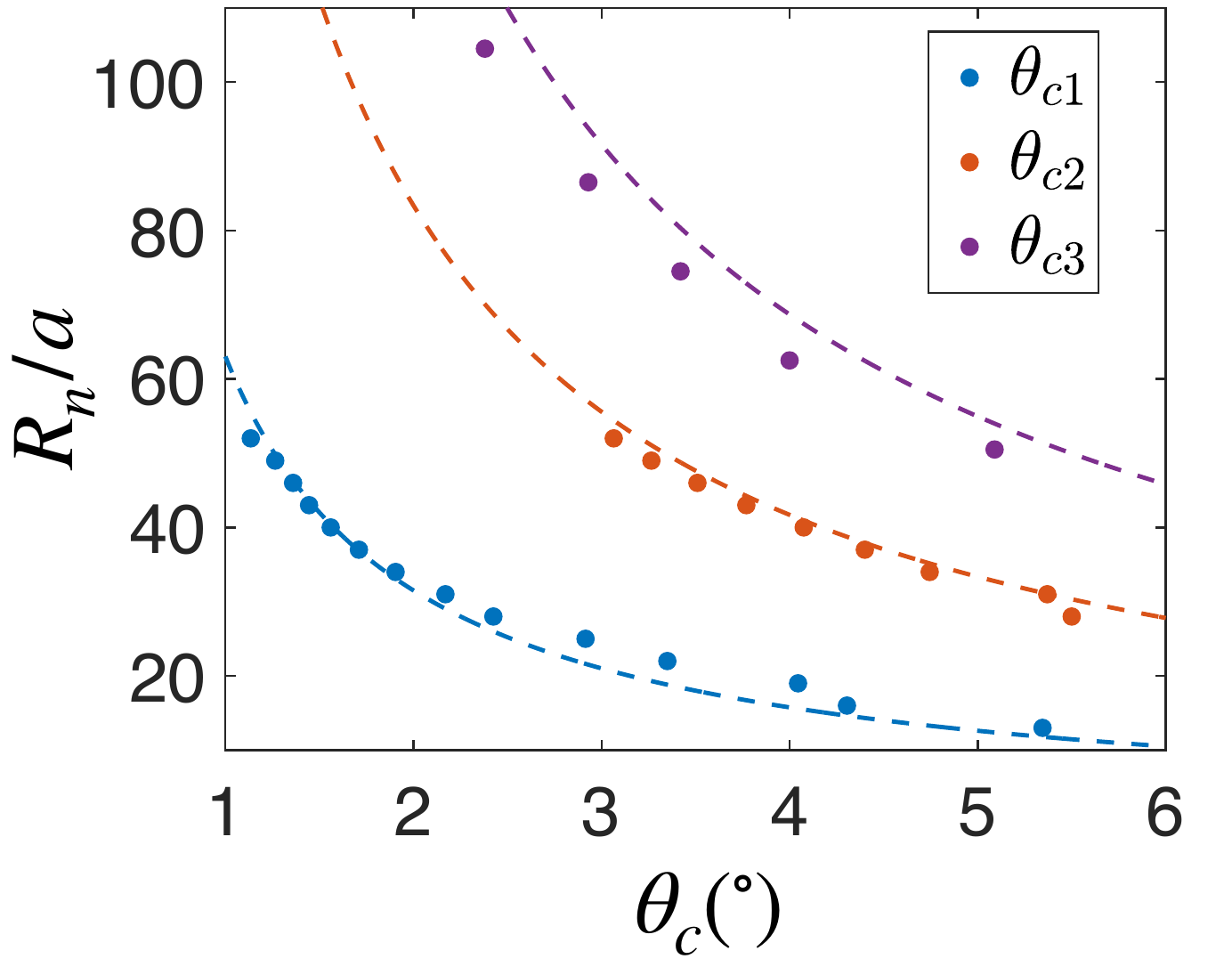}
\caption{Scaling relation of the radius of the top layer disc $R$ with respect to the resonance angles $\theta_c$ for the first, second and third resonant peaks. Here the Fermi energy is $E_F=2.0$ eV, $V_{d}=5$ eV. For the first and second resonance curves, the calculations were done for systems with $W=2R$, while for the third resonance curve, $W$ was equal to $R$ to reduce the computational difficulty. Each data (see the dots) was obtained by reading from the second derivative of the 9-th polynominal fitted curves from the averaged conductance curves. The dashed lines are the theoretically estimated scaling relation between $R$ and $\theta_c$. Here a modification factor $\zeta = 1.1$ has been used to account for the overfilling of the moir{\'e} pattern inside the top disc.  }
\label{fig: ScalingRelation}
\end{figure}

The period of the moir{\'e} pattern of the TBG is defined as the distance between any two adjacent AA stacking region, or unit-moir{\'e} supercells, and is calculated to be $L = \sqrt{3}a/(2 \sin{\theta/2})$. The radius of a circle emcompassing exactly one hexagonal moir{\'e} supercell is calculated to be $R_{n} = \sqrt{S_n}L/{\sqrt{3}}  =  \sqrt{S_n}a / (2 \sin{\theta_{cn}/2})$ for the $n$-th resonance angle $\theta_{cn}$. To guarantee the full encirclement of the whole hexagonal moir{\'e} supercell by the top disc, the radius of the circle in real transport process should be slightly larger than $R_n$ and thus we here consider a modification factor $\zeta$: $R_{n} \rightarrow \zeta R_{n}$. In Fig. \ref{fig: ScalingRelation} we show the scaling relation of the $n$-th resonance angle $\theta_{cn}$ with the radius $R_n$ of the top disc which are shown with discrete dots as read from the second derivative of the 9-th polynominal fitted curves from the averaged transmission coefficient $T_A$. To make a comparison, we also show the theoretically estimated scaling relation $R_{n}(\theta)$ with dashed lines. The modification factor $\zeta$ has been chosen to be 1.1. One can see good agreement between the numerically calculated dots from quantum transport and the theoretically estimated ones, which further validates our explanation on the resonant peaks in quantum transport based on the geometric moir{\'e} patterns. 

\section{Discussion and conclusions \label{sec: Conclusion}}
The top-layer disorder configuration used in our calculations is crucial for generating the resonance tunneling through the TBG, because the system we are confronting is a mesoscopic one, where quantum fluctuation as a result of the interference effect would mask the QD effect which exists naturally in TBG systems. Even when the transport system is clean, the TBG region which inherently contains nonuniform interlayer coupling and irregular boundary against the leads made of monolayer graphene, works as a chaotic system that transmits electrons with random probabilities due to the phase randomness. The quantum fluctuation disappears only when the rotation angle is large ($\theta > 15^{\circ}$) which decouples the bilayer system. Disorder, however, after enough ensemble averages, removes the phase randomness and smears out the fluctuation, and finally, unravels the hidden geometric effect by showing the resonant tunneling arising from the moir{\'e} structure. Here we also want to emphasize that the resonant tunneling as arising from the formation of the moir{\'e} pattern is newly reported in this paper which provides a new perspective into disordered systems. 

The geometric resonance reported here should be observable in experiments since it happens on a mesoscopic scale and does not require a periodic moir{\'e} structure. The robustness of the resonant peak against the disorder strength, the Fermi energy, and the shape distortion of the TBG region makes the experimental observation expedient with no requirement for subtle control of those parameters. There are two points to be emphasized for experimental observations: (1) one should keep the area of the TBG region invariant upon rotation to avoid other disturbances on the transport results, and (2) the disorder effect should asymmetrically exists mainly on the top layer. If disorder exists on both layers, the system becomes a trivial mesoscopic conductor with no such resonance tunneling effect (see Appendix D). To make ensemble average, one can fix one disorder configuration by chemical doping or applying a top gate with random potential, and then scan the Fermi energy or an external magnetic field (the magnetic field should be small enough to avoid localization or anti-localization effect). Actually thousands of ensemble averages are not necessary as we have done in this paper: tens of ensemble averages should be enough to see the resonant phenomenon (see Appendix C). 

In conclusion, we investigate quantum transport through a TBG system on the mesoscopic scale, where the TBG system consists of a disordered top layer graphene disc and a clean bottom graphene nanoribboon. We find that with strong disorder, the averaged transmission through the TBG system shows a sequence of resonant peaks with respect to the rotation angle, with the resonance angles $\theta_c$ being robust against the disorder strength, the Fermi energy, and the shape distortion of the central TBG region. We plot the moir{\'e} patterns inside the TBG region at the resonance angles and find that the resonance happens when the TBG boundary encompasses one giant hexagonal moir{\'e} supercell, and  thus has a purely geometric origin. We explain this geometric resonance in terms of the averaged DOSs inside the moir{\'e} pattern which are localized at the center of  AA stacking region while minimizes at the AB/BA stacking regions. Finally the scaling relation of the size of the TBG with respect to the resonance angles $\theta_c$ is also given, which shows agreement with the theoretical analysis based on the moir{\'e} structure. The results reported here provide a new way to control the conductance in twisted moir{\'e} systems by the rotation angle and should be experimentally observable in a two-terminal mesoscopic system where a nonsymmetric distribution of defects is possible due to adatoms, admolecules or substrate effect, etc\cite{Namarvar2020Transport}. 

\textit{Note added}: After finishing this work, we
become aware of a similar work in \cite{Moles2023}.

\section*{acknowledgements}
We thank L.-W. Fu and W.-L. Zeng for the technical supports on using the Qlab server. We thank X.-C. Xie, J.-H. Gao, and Q.-F. Sun for helpful discussions on the draft. The work is supported by National Key Basic Research Program of China (No. 2020YFB0204800), the National Science Foundation of China (Grant No. 12204432), and Key Research Projects of Zhejiang Lab (Nos. 2021PB0AC01 and 2021PB0AC02).

\section*{Appendix A: Algorithm in calculating the transmission coefficients through the TBG system}
\def\theequation{A\arabic{equation}}
\setcounter{equation}{0}

\begin{figure}
\includegraphics[width=8.0cm, clip=]{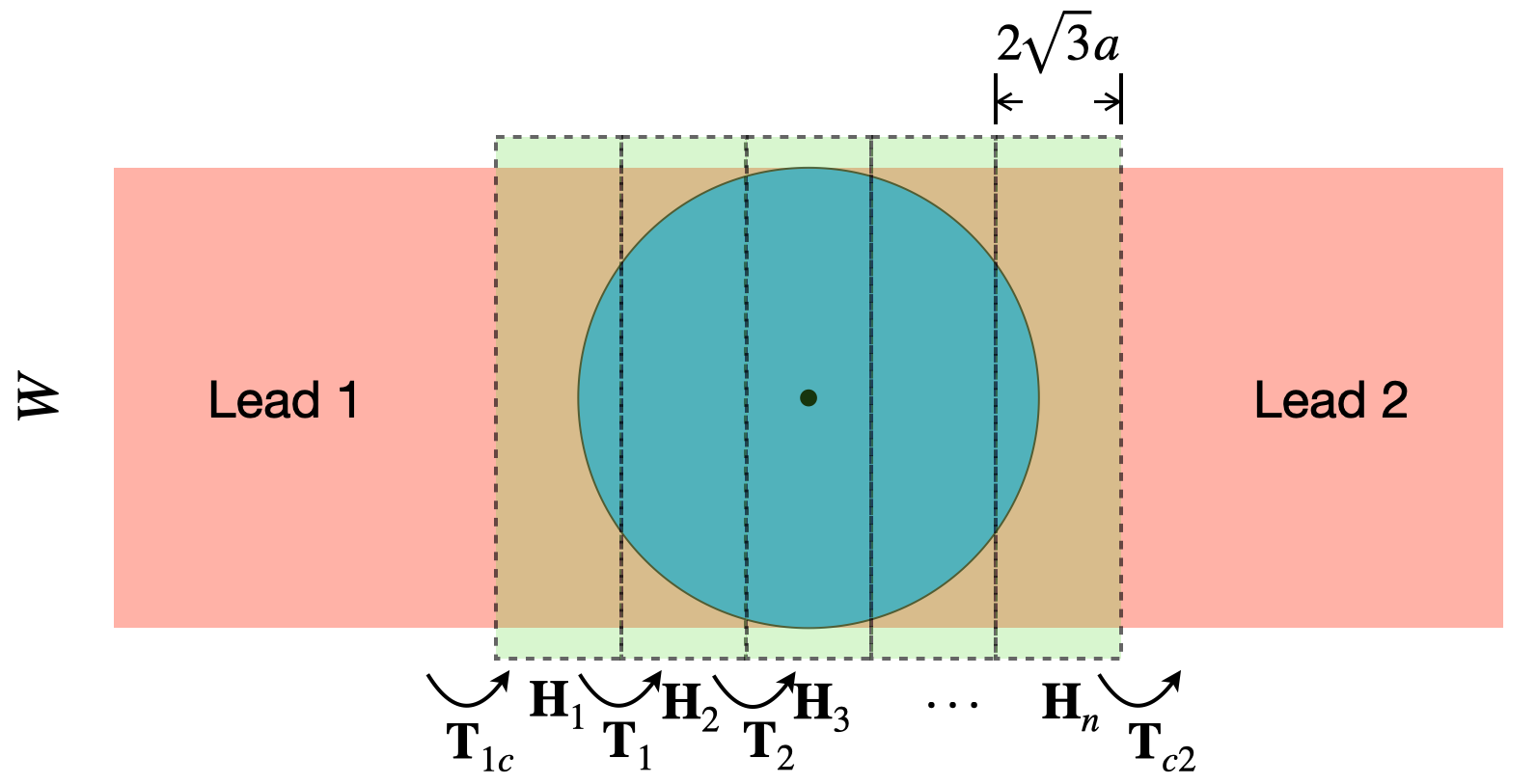}
\caption{Schematic diagram of the two-terminal TBG transport system, showing the algorithm of calculating the Green's function ${\bf G}^r_c$ of the central region (see the green region composed with dashed blocks). Here central region is divided uniformly into several blocks with width $2 \sqrt{3}a$. Since we have already set the hopping boundary to be $2\sqrt{3}a$ in the tight-binding model, only the adjacent blocks contribute to nonzero hopping matrices ${\bf T}_i$ )(here $i$ denotes the two adjacent blocks $i$ and $i+1$. The matrix for the $i$-th block can be obtained by numerating the coordinates of each carbon atoms within it. The hopping matrices ${\bf T}_{1c}$ and ${\bf T}_{c2}$ denote the hopping matrices between the central region and the leads L(R).  }
\label{fig: Iterative_Algorithm}
\end{figure}

To get the retarded Green's function ${\bf G}^r_c$ of the central region numerically as can be seen in Sec. \ref{sec: Model and Hamiltonian}, we used an iterative Green's function method. The detailed algorithm can be seen here: 1) Get all the coordinates of all bottom sites within the central region, the on-site terms and hopping elements of the bottom layer; 2) Get all the coordinates of all bottom layer within the circle (overlapping region); 3) Rotate the coordinates of all bottom sites within the circle to get the coordinates of all top sites; 4) Sort all the top coordinates of the top layer by ascending $x$-coordinate; 5) Divide the central region into several blocks with width $2\sqrt{3} a$ (see Fig. \ref{fig: Iterative_Algorithm}), so that only the nearest blocks have overlapping hopping integrals. 6) Write down the Hamiltonian ${\bf H}_i$ of the $i$-th block by calculating the on-site terms and the hopping terms between any two carbon atoms. 7) Write down the hopping matrix ${\bf T}_{i}$ between two adjacent blocks $i$ and $i+1$, and the hopping matrices ${\bf T}_{1c}$ and ${\bf T}_{c2}$ between the central region and two leads. 8) After, use the iterative Green's function for the one-dimensional prototype to get the Green's function ${\bf G}^r_c$.

\section*{Appendix B: Periodicity of the transmission coefficient $T(E)$ through the TBG region}
\def\theequation{B\arabic{equation}}
\setcounter{equation}{0}

\begin{figure}
\includegraphics[width=8.7cm, clip=]{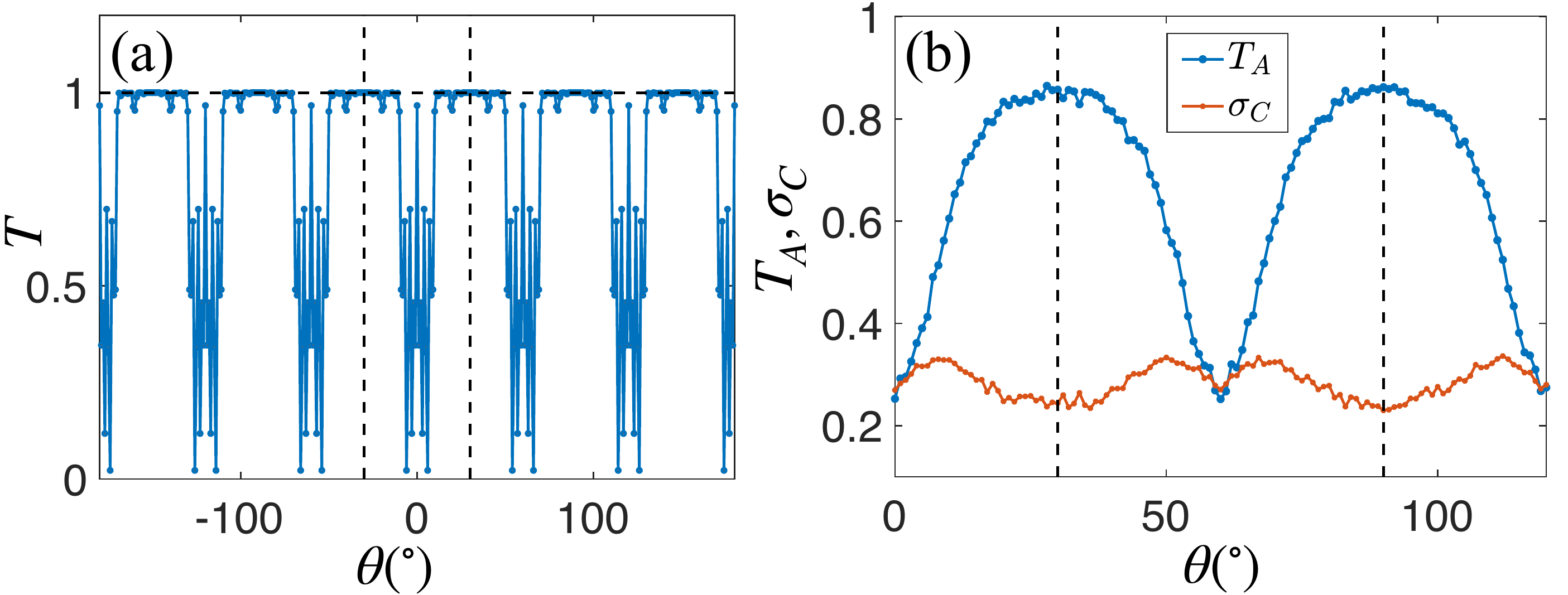}
\caption{(a) Transmission coefficient $T$ through the two-terminal TBG system as a function of the rotation angle $\theta$ without disorder. The width $N=50$, the incident energy $E=0.85$ eV, the interlayer distance $d=a_I$, and the radius of the top disc is $R=W/2$. The two dashed lines denote the positions of $\theta = \mp 30^{\circ}$.  (b) Averaged transmission coefficient $T_{A}$ and conductance fluctuation $\sigma_C$ as a function of the rotation angle $\theta$. Here the the width of the nanoribbon $N=50$, the interlayer distance $d=a_I$, the incident energy is $E=0.85$ eV, the radius of the top disc is $R=W/2$, and the disorder strength $V_{d}=8$ eV. Here the disorder exists only on the top layer. }
\label{fig: transmission_N50_AllRange}
\end{figure}

The peridocity of the transmission $T(T_{A})$ through the TBG region without (with) disorder is shown in Fig. \ref{fig: transmission_N50_AllRange}. The transmission shows a peridocity of $60^{\circ}$ as a result of the circular shape of the central TBG region. The transmission also shows a mirror symmetry about $\theta= \pm 30 ^{\circ}$ [see the black dashed lines in Fig. \ref{fig: transmission_N50_AllRange} (a) and (b)] due to the same structure for $\theta$ and $-\theta$, so it is quite resonable to consider only the rotation angle within $[0, 30^{\circ}]$. Besides the oscillation of the averaged transmission coefficient $T_A$ in Fig. \ref{fig: transmission_N50_AllRange}(b), the conductance fluctuation $\sigma_C$ also shows oscillations with respect to $\theta$ which minimizes at $\theta = 30 ^{\circ}$ and peaks at $\theta = 9 ^{\circ}$.

\section*{Appendix C: Convergence of the transmission coefficients and conductance fluctuation}
\def\theequation{C\arabic{equation}}
\setcounter{equation}{0}

\begin{figure}
\includegraphics[width=8.7cm, clip=]{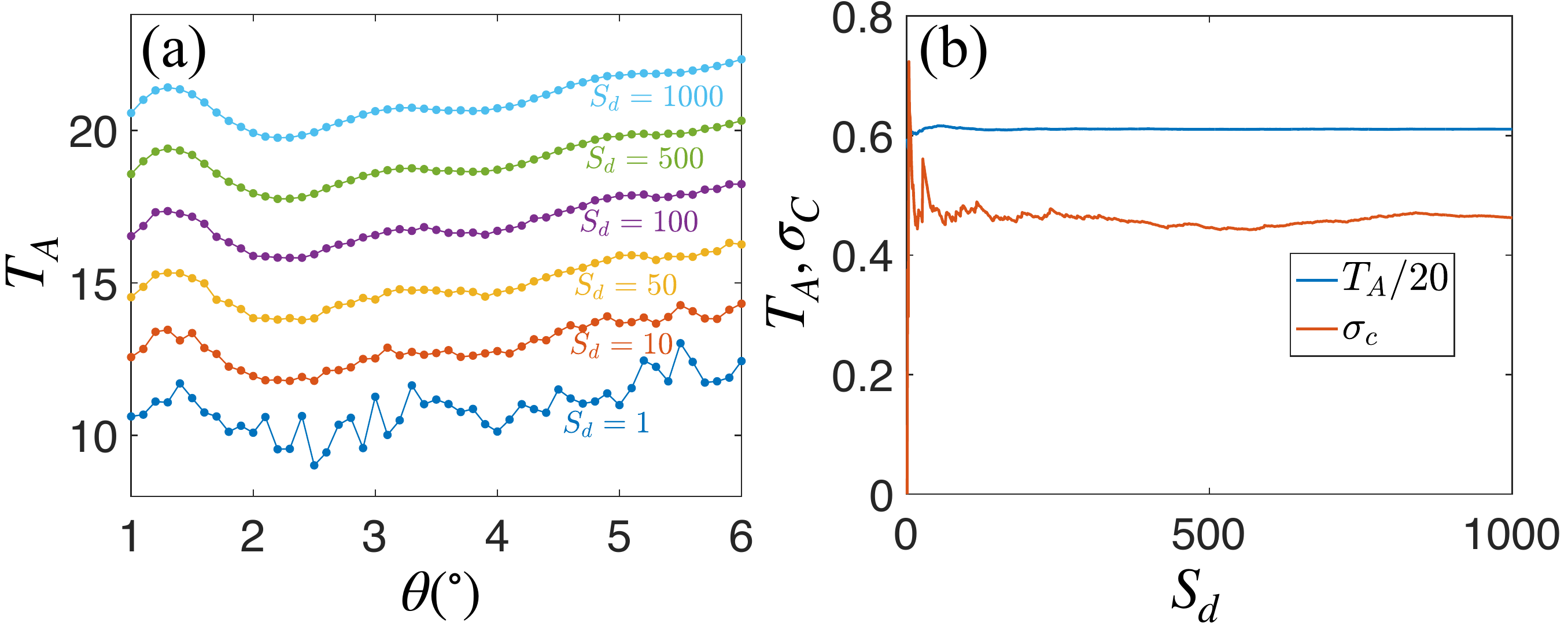}
\caption{(a) The averaged transmission coefficient $T_{A}$ as a function of the rotation angle $\theta$ for different ensemble average times $S_d$. (b) Convergence of the averaged transmission coefficient $T_{A}$ and the conductance fluctuation $\sigma_c$ at angle $\theta=6^{\circ}$ as varying the ensemble averge times $S_d$. Here to show them in the same scale, we have divided $T_{A}$ by 20. For (a) and (b) we choose $N=70, E_F = 2 {\rm eV}, V_{d} = 5 {\rm eV}$ and $R= W/2$. The disorder exists only on the top layer.  }
\label{fig: TestConvergence}
\end{figure}

To test the convergence of the averaged transmission $T_{A}$ in the main text and the conductance fluctuation $\sigma_c$ calculated from a fixed number of ensemble averages, we first plot the $T_{A}$ curves with respect to the rotation angle $\theta$ for the disorder average times $S_d$ = 1, 10, 50, 100, 500, 1000 as can be seen in Fig. \ref{fig: TestConvergence}(a). Here we have choose the TBG region in a disc shape with $R = W/2$. We find that, for one disorder figuration, the resonant peaks can hardly be distinguished due to the strong fluctuation of the transmission as varying $\theta$. After making 10 ensemble averages, the first resonant peak can be distinguished. With increasing the averge times, the transmission curves become more smooth, and higher resonant peaks can be clearly distinguished. In Fig. \ref{fig: TestConvergence}(b), we choose the rotaion angle $\theta=6^{\circ}$, and give the $S_d$-dependence of the averaged transmission $T_{A}$ and conductance fluctuation $\sigma_c$. Here to show both of them in the same window, we have divided $T_{A}$ by a factor 20. We see that, the avearged transmission $T_{A}$ converges quickly after making 100 ensemble averages. However, the conductance fluctuation converges well only after making around 1000 times of averages. To gurantee the ergodicity of our ensemble, we thus choose $S_d=1000$ in all disorder calculations in the main text. However, to get the (or to measure) the resonant peaks in the transmission curves, tens of ensemble averages should be enough, which makes the experimental verification of our results quite feasible. \\

\section*{Appendix D: Avereged transmission coefficients and conductance fluctuation with both-layer disorder}
\def\theequation{D\arabic{equation}}
\setcounter{equation}{0}

\begin{figure}
\includegraphics[width=8.5cm, clip=]{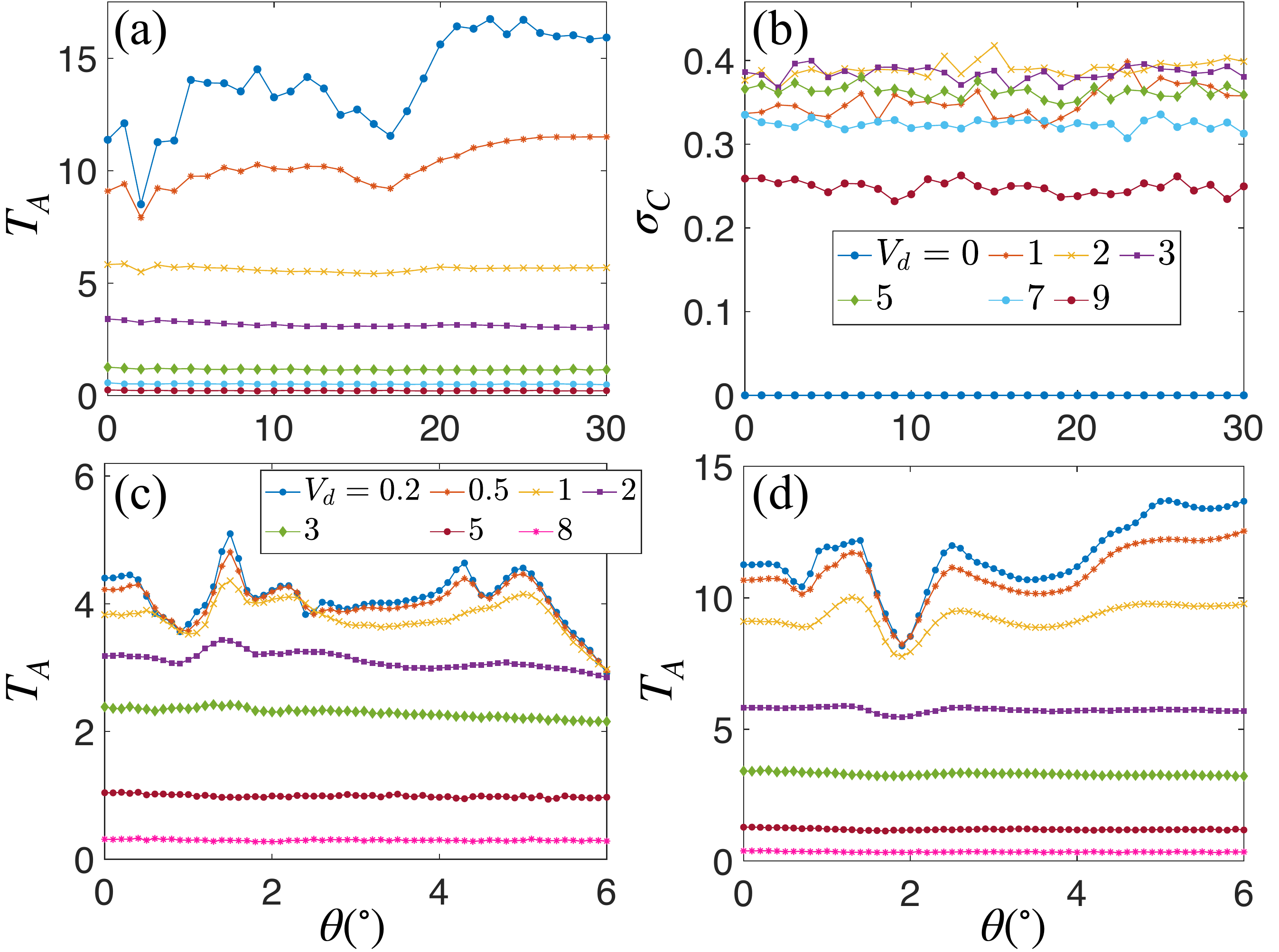}
\caption{The calculations with both-layer disorder. (a) and (b): The averaged transmission coefficient $T_{A}$, and the conductance fluctuation $\sigma_{C}$ (in units of $e^2/h$) as a function of the rotation angle $\theta$ under different disorder strength $V_d$. Here the Fermi energy $E_F=2$ eV. (a) and (b) share the same legend. (c) and (d): The averaged transmission coefficient $T_A$ as a function of rotation angle $\theta$ for $E_F=1.36$ eV and $E_F=2$ eV, respectively. For (a-d)  $N=50$, $d = a_I$, $R= W/2$. The disorder is averaged for 1000 times. }
\label{fig: TestConvergence}
\end{figure}

In Fig. \ref{fig: TestConvergence} we show the calculations on the transport through the TBG device with both-layer disorder. Here the central TBG region is confined within the disc region as we have considered in Sec. \ref{sec: Transport Results with circular boundary}. In Fig. \ref{fig: TestConvergence} (a, b) we fixed the Fermi energy $E_F = 2$ eV, and show the averged transmission coefficient $T_A$ and the conductance fluctuation $\sigma_C$ as a function of the rotation angle $\theta$ under different both-layer disorder strength $V_d$. Here the same curve color is used for the same $V_d$. We see that, for weak disorder $V_d < 1$ eV, the transmission gets an enhancement after $\theta > 15 ^{\circ}$ due to the decoupling regime of the TBG and shows dependence (fluctuation) with the rotation angle. However, after the disorder strength is increased, the angle dependence(fluctuation) is smeared out and the averaged transmission coefficient $T_A$ shows a constant regardless of the rotation angle, consistent with the calculation in Ref.\cite{Namarvar2020Transport}. The conductance fluctuation $\sigma_C$ also shows no dependence on the rotation angle. Besides, the transmission gets decreased and approaches zero as $V_d$ increases, which indicates that strong both-layer disorder tends to localize the electrons inside the TBG region, which is quite different from the top-layer disorder case. For the intermediate disorder strength $1 \leq V_d \leq 5$ eV, the conductance fluctuation $\sigma_C$ is very close to the universal value of mesoscopic conductance fluctuation $0.4$, while for strong disorder $\sigma_C$ gets decreased as the transport goes into the localisation regime. These behaviors all indicate that the transport throuth a TBG system is quite similar to the normal 2D systems when both-layer disorder is considered. 

To further validate that there is no such geometric resonance in the both-layer disorder case, in Fig. \ref{fig: TestConvergence}(c, d) we show the transport results within a small range of rotation angle ($\theta \in [0, 6^{\circ}]$) by scanning the disorder strength $V_d$. We show that, disorder has only two effects on the averaged transmission coefficient $T_A$: (1) smears out the fluctuation with respect to the rotation angle, and (2) suppresses the transmission. No such resonant peaks are observed in Fig. \ref{fig: TestConvergence}(c, d) around the two moir{\'e} resonance angles $\theta_{c1}$ and $\theta_{c2}$.


\begin{thebibliography}{10}
\bibitem{Footnote1} Actually the parameter $\gamma_0 =3.7 $eV yields a Fermi velocity for the monolayer graphene $v_F \approx 1.1 \times 10^{6} \rm m/s$ which fits the experimental data well. In our calculations, we choose $\gamma_0 = 2.7 $eV which shows a small derivation from the experiments but still holds good agreement with the DFT calculations.
\bibitem{Footnote_localization} The localization length is calculated by considering a graphene nanoribbon with a rectangular disordered region with the Green's function method.
\bibitem{Footnote2} We take the function $f(x) = a x + \sin{x}$ as an example. The positions of the peaks locate at $a = \cos{x}$ which yields $x_c = \arccos{a}$ and moves leftward as $a$ increases. We ascribe the moving of the first resonant peak to the same reason here.
\bibitem{Footnote3} The higher Fermi level crosses more transmission channels from the leads so that the density of states within the TBG region beome more, and as a result, the resonant transmission mediated by the quantum dot structure formed naturally within the TBG moir{\'e} pattern becomes prominent.
\end{thebibliography}
\end{document}